\documentclass[prl,aps,twocolumn,showpacs]{revtex4}
\usepackage{graphicx}
\usepackage{amsmath}
\usepackage{bm}

\newcommand{\be}{\begin{equation}}
\newcommand{\ee}{\end{equation}}
\newcommand{\bea}{\begin{eqnarray}}
\newcommand{\eea}{\end{eqnarray}}
\newcommand{\bei}{\begin{itemize}}
\newcommand{\eei}{\end{itemize}}

\begin{document}

\title{Measuring the Superfluid Fraction of an Ultracold Atomic Gas}

\author{Nigel R. Cooper and Zoran Hadzibabic}
\affiliation{Cavendish Laboratory, University of Cambridge, J.~J.~Thomson Ave., Cambridge CB3~0HE, U.K.}

\begin{abstract}

  We propose a method to measure the superfluid fraction of an
  atomic gas.  The method involves the use of a vector potential
  generated by optical beams with non-zero angular momentum to
  simulate uniform rotation. The induced change in angular momentum
  of the atomic gas can be measured spectroscopically. This allows a
direct determination of the superfluid fraction.

\end{abstract}
\date{26 October, 2009}


\maketitle

Central to the understanding of the physics of degenerate Bose gases
are the concepts of Bose-Einstein condensation and superfluidity~\cite{griffinss,leggettsuperfluidity}.
Bose-Einstein condensation refers to the macroscopic occupation of a
single quantum state. Superfluidity refers to a set of fascinating
hydrodynamic phenomena, notably persistent (dissipationless)
flow.

Both phenomena admit clear quantitative definitions, allowing a Bose
gas to be characterised by ``condensate'' and ``superfluid'' fractions~\cite{griffinss,yangodlro}.
These two quantities 
in general take very different values.
A gas of non-interacting bosons at low temperature forms
a Bose-Einstein condensate (BEC), but is not superfluid.
In the low-temperature limit, liquid $^4$He is both a BEC and a
superfluid, but the condensate and superfluid fractions are markedly
different, believed to be $\sim 10\%$
and $100\%$, respectively~\cite{griffinss}. In 
2D systems
the superfluid fraction can be non-zero even if the condensate fraction vanishes.

For ultra-cold atomic Bose gases, the condensate fraction is readily
measured through the mapping of occupation numbers in momentum space
to real space by expansion imaging~\cite{bec95}.  Characteristic
signatures of superfluidity have been observed in atomic gases,
notably dissipationless flow~\cite{ramancriticalvelocity,phillipstoroidal},
and the formation of quantized vortices
in rotating gases~\cite{madison}. However, there has been no quantitative measurement of the
superfluid fraction.  Such a measurement is crucial for the
investigation of some of the most interesting properties of
interacting Bose gases~\cite{blochdz}: strong interactions can lead to condensate
depletion without loss of superfluid fraction;
the Kosterlitz-Thouless phase transition
in a quasi-2D geometry is
manifest by a universal jump of the superfluid
density~\cite{KTjumpbishopreppy}.

In this Letter, we describe how the superfluid fraction of an atomic gas
can be measured using a light-induced vector potential~\cite{blochdz}. Our method
is closely analogous to the classic experimental method of
Andronikashvili~\cite{andron}.
There, liquid helium is put in contact with a rotating object: the normal fluid
picks up non-zero angular momentum, while the
superfluid acquires no angular momentum. A measurement of the
angular momentum of the fluid then allows a determination of the
superfluid fraction.  Here, we consider the use of an optically
induced vector potential to simulate  uniform rotation~\cite{blochdz}.
We show how  spectroscopy can be used to measure the net change in angular momentum of the fluid,
and hence the superfluid fraction.
Recently, measurement of the superfluid fraction from the density profile of  a rotating gas
was also proposed~\cite{ho-phasediagram}.
Throughout the paper, we consider a gas of identical bosons, but the
method can be extended to other situations, such as superfluidity of
paired fermions.

The definition of the superfluid fraction was expressed in a form
suitable for our purposes by Leggett~\cite{leggettssleggettt0}.  It
applies to a fluid contained in a ring-shaped vessel with a radius $R$ that is
large compared to its transverse
dimensions, so that the classical moment of inertia is $I_{\rm cl} = NM
R^2$ for $N$ atoms of mass $M$.  We start by adopting this
assumption of geometry, but this will be relaxed at the end of
the paper. The walls of the vessel are taken to rotate with angular
velocity $\omega$, and the fluid allowed to come to thermal
equilibrium. Under these conditions, the superfluid fraction
is~\cite{leggettssleggettt0}
\be \frac{\rho_s}{\rho} \equiv 1 - \lim_{ \omega\to 0} \left(\frac{\langle L
    \rangle}{I_{\rm cl} \omega}\right)\,,
\label{eq:sfdefinition}
\ee
where $\langle L\rangle$ is the average angular momentum of the fluid.
A normal fluid will come to rest in the frame rotating with the
walls, so that $\langle L \rangle = I_{\rm cl}\omega$ and $\rho_s/\rho=0$.  A
(perfect) superfluid is unaffected by the rotating walls, so $\langle L\rangle =0$
and $\rho_s/\rho=1$.

When the fluid is in equilibrium with the rotating walls
it is described  by the equilibrium
  density matrix defined by the Hamiltonian in the rotating frame~\cite{LLStatMech}
\be
H_{\rm rot} = H - \bm{L}\cdot\bm{\omega}\,.
\label{eq:hrot}
\ee
Here, $H$ and $\bm{L}$ are the Hamiltonian and the total angular momentum
in the {laboratory frame}.
We shall show how a Hamiltonian of the form
(\ref{eq:hrot}) can be achieved for an atomic gas, and how the
resulting average angular momentum $\langle L\rangle$ can be measured
so that (\ref{eq:sfdefinition}) can be directly applied.

In the ring geometry,
the kinetic energy in (\ref{eq:hrot}) can be written
\be
\label{eq:angham}
\frac{\bm{p}^2}{2M} - \bm{r}\times\bm{p}\cdot\bm{\omega}
 =
\frac{{\bm p}_\perp^2}{2M}
+\frac{\hbar^2}{2MR^2} \ell^2  - \hbar\omega \ell \, ,
\ee
where $\bm{p}_\perp$ is the momentum in directions perpendicular to
the azimuthal direction, and $\ell$ is
the angular momentum in units of $\hbar$ (therefore quantized to integer values). The rotation shifts the energy minimum in the angular momentum to
\be
\ell^* = \frac{MR^2\omega}{\hbar} = \frac{I_{\rm cl}\omega}{N\hbar}\,.
\label{eq:lstaromega}
\ee
This shift can be viewed as an azimuthal vector
potential corresponding to a non-zero flux threading the ring.

A shift in the dispersion relation can be
achieved by the use of two-photon Raman transitions to imprint vector potentials~\cite{blochdz}. This was recently implemented~\cite{lin:130401} using two counter-propagating laser beams to couple states $m=-1,0,1$ of the $F=1$ hyperfine levels of
$^{87}$Rb. The two-photon processes
lead to a linear vector potential directed along the axis
of the lasers~\cite{lin:130401,spielmanpra}.

To generate an azimuthal vector potential, we
consider two Laguerre-Gauss (L-G) beams~\cite{laguerregauss} with different
orbital angular momenta, co-propagating in the direction perpendicular to the toroidal trap.
In this way, a two-photon transition imparts negligible linear momentum to the atoms, but
a non-zero angular momentum, $\pm \Delta \ell$, where $\Delta \ell$ is the
difference in the orbital angular momenta of the two beams.
For a 3-level system~\cite{lin:130401,spielmanpra} this leads to an effective  Hamiltonian
$$
\left(\begin{array}{ccc}
    \frac{\hbar}{2MR^2} (\ell+\Delta \ell)^2 -\delta & \Omega_R/2 & 0\\
    \Omega_R/2  & \frac{\hbar}{2MR^2} \ell^2 -\epsilon & \Omega_R/2 \\
    0 & \Omega_R/2 & \frac{\hbar}{2MR^2} (\ell-\Delta \ell)^2
    +\delta \end{array}\right)
$$ which is a matrix in the hyperfine
states $m=-1, 0, +1$.
$\Omega_R$ is the two-photon Rabi frequency,
$\delta= g_F \mu_{\rm B} B/\hbar$ is the detuning of the lasers
from the Raman resonance set by the Zeeman effect of a (uniform) magnetic field $B$,
and $\epsilon$ accounts for the quadratic Zeeman effect.

\begin{figure}[ttp]
\includegraphics[width=0.9\columnwidth]{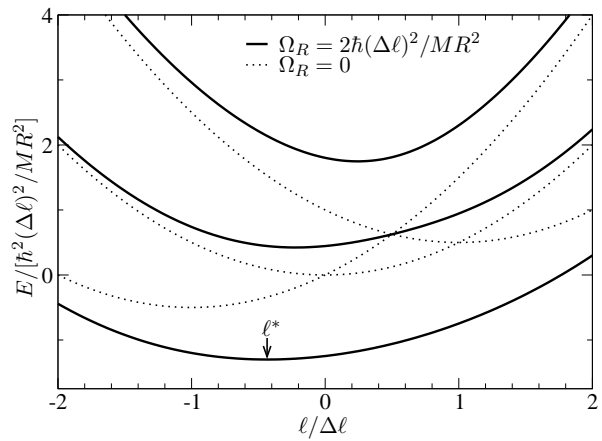}
\caption{ \label{fig:curves}
Energy levels for the angular motion of an atom
under the influence
of two-photon Raman coupling via beams with orbital angular momentum difference $\Delta \ell$.
The atoms move in a trap of radius $R$, the detuning is $\delta = 0.5 \hbar (\Delta \ell)^2/MR^2$, $\epsilon=0$, and Rabi frequencies of $\Omega_R=0$ (dotted line) and
$\Omega_R=2\hbar (\Delta \ell)^2/MR^2$ (solid lines) are shown. The lowest band has its minimum displaced to a nonzero angular momentum, $\ell^*$, equivalent to the effect of an azimuthal vector potential.
(Note that the smooth curves should be viewed as interpolations between
the allowed integer values of $\ell$.)
}
\end{figure}

The energy eigenvalues for the angular motion are illustrated in
Fig.~\ref{fig:curves}.  The
lowest energy band has a minimum at a non-zero angular
momentum $\ell^*$.
To derive analytic expressions for the energies and
wavefunctions in this band, we
consider the limit of large Rabi frequency
and develop a
perturbation theory in $1/\Omega_R$.
We parameterise the resulting single-particle energy as
\be E = E_0
+ \frac{\hbar^2}{M^* R^2} \left(\frac{\ell^2}{2}- \ell \; \ell^*\right) \,.
\label{eq:eeff}
\ee
The energy $E_0$ is a global shift that depends on all parameters of
the optical field.
A slight increase in the effective mass for the azimuthal motion is given by
\be
 M^* = M \left(1 +
  \frac{\sqrt{2}\hbar \Delta \ell^2}{M R^2 \Omega_R}\right)  +   {\cal O}(1/\Omega_R^2) \,.
\ee
The most important effect for our purposes is the shift in the minimum of the dispersion curve to
\be
\ell^* = -\sqrt{2} \frac{\delta}{\Omega_R} \Delta \ell +
 {\cal O}(1/\Omega_R^2)\,.
\label{eq:lstar}
\ee
This can be viewed as the introduction of an effective vector potential.
In order to clarify the physical consequences of $\ell^*$,
note that, for an atom with angular momentum $\ell$,
the angular velocity when the light is on is
\be
\label{eq:wlight}\omega_{\rm light}  \equiv   \frac{1}{\hbar} \frac{d E}{d \ell} = \frac{\hbar}{M^* R^2} \left(\ell - \ell^*\right) \,.
\ee
Thus, for given $\ell$, the most significant effect of the light
is to cause a
 {\it constant shift in the angular velocity} by ${\hbar \ell^*}/{M^* R^2}$.
In
analogy with (\ref{eq:angham},~\ref{eq:lstaromega}),
it is as if
the optical field causes the laboratory
frame to behave as a  frame of reference that is rotating with angular
frequency
\be
\omega_{\rm eff} \equiv \frac{\hbar \ell^*}{M^* R^2}\,.
\label{eq:weff}
\ee

The lowest band (\ref{eq:eeff}) plays the role of $H_{\rm rot}$ (\ref{eq:hrot}), with
an effective rotation rate (\ref{eq:weff}) that can be tuned by the parameters of the optical field.
Provided the splitting of the bands is large compared
to the chemical potential, $\Omega_R \gtrsim \mu/\hbar$, all atoms are
restricted to this lowest band.
Then, if the position of its
minimum (\ref{eq:lstar}) is varied sufficiently slowly that the fluid has time to come
to equilibrium for the new $\ell^*$, a clear distinction appears
between normal and superfluid components: the normal fluid
will relax, and pick up a non-zero angular momentum;
the superfluid will {\it not} relax, but will retain vanishing average
angular momentum. This distinction provides the {definition} of the
superfluid fraction (\ref{eq:sfdefinition}).

This behaviour of the atomic gas in a toroidal trap with
an azimuthal vector potential is in marked contrast to the case of a
linear vector potential~\cite{lin:130401}. There, both superfluid and
normal components must come to rest in the laboratory frame,
consistent with the observations in Ref.~\cite{lin:130401}.  Here,
when the normal fluid comes to equilibrium at a non-zero $\ell^*$,
it is at rest in the laboratory frame. However, the superfluid is rotating,
as follows from (\ref{eq:wlight}). The azimuthal vector potential causes a
steady superfluid flow around the ring-shaped trap.
Note that for the normal fluid to come to equilibrium with the new $\ell^*$, it must change its angular
momentum. Therefore, the trap must {\it not} be perfectly rotationally
symmetric. ({No fluid can come to equilibrium in a
rotating container if the walls of the container are perfectly
smooth.})
It is an important practical feature that there is no requirement for the trap to
have perfect rotational symmetry.

In the Andronikashvili experiment~\cite{andron}
a torsional oscillator is used to measure the moment
of inertia of the fluid coupled to the oscillator, $I$.
A frequency shift arises from the fluid's contribution to the energy, which in the rotating frame is
$\langle H_{\rm rot}\rangle = -\frac{1}{2} I \omega^2$.
  Here, when the light is on
 there is a contribution to the energy
 of $-\frac{1}{2} I \omega_{\rm eff}^2$. One can envisage various ways in which an
oscillator can modulate $\omega_{\rm eff}$ (\ref{eq:weff}) and therefore
experience a frequency shift related to $I$.
For example, given that $\omega_{\rm eff} \propto \ell^* \propto B$,
 the coil that generates
the Zeeman field $B$ experiences the moment of inertia as a reduction in
its inductance; this  could
 appear as a shift in the resonant frequency of an
electrical circuit containing the coil.
However, the {\it total} energy $\frac{1}{2} I (\hbar \Delta \ell/M^* R^2)^2$
is very small ($\simeq 0.1 \mu\mbox{eV}$)  making the signal small compared to typical sensitivities of current micro-mechanical or electrical oscillators.

A key element of our proposal is that, with the above light-induced vector
potential, one can use {\it spectroscopic} methods to determine the
average angular momentum $\langle L\rangle$. The wavefunction in the
lowest band is a linear superposition of the three hyperfine levels
$|\psi\rangle  \equiv \sum_{m=-1,0,1} \psi_m |m\rangle$
with amplitudes
$\{\psi_m\}$ which vary with
 $\ell$.
A perturbative
analysis shows that there are
equal and opposite corrections to $|\psi_{\pm 1}|^2$ which depend {\it linearly}
on $\ell$.   Thus,  $\langle L \rangle$
can be obtained from a measurement of the difference
in the number
of particles in the states $m=\pm 1$.
Using the Taylor expansion
\be
\label{eq:taylor}
|\psi_{-1}|^2 - |\psi_{1}|^2 \equiv \Delta p_0 + \Delta p' \ell +{\cal O}(\ell^2)
\ee
we can write
\be
\frac{\langle L\rangle}{\hbar N} \equiv \frac{\sum_\ell \langle n_\ell\rangle \ell}{\sum_\ell \langle n_\ell\rangle}
= \frac{\Delta p - \Delta p_0}{\Delta p'}
\label{eq:lavgen}
\ee
where we define
the fractional imbalance
\be
\label{eq:deltap}
\Delta p \equiv \frac{N_{-1} - N_{1}}{N} =
\frac{\sum_\ell \langle n_\ell\rangle \left[|\psi_{-1}|^2 - |\psi_{1}|^2\right]}{ \sum_\ell \langle n_\ell\rangle}
\,,
\ee
with $\langle n_\ell\rangle$ the average number of particles with angular momentum $\ell$.
Inserting (\ref{eq:lavgen}) in (\ref{eq:sfdefinition}) one finds
\be
\frac{\rho_s}{\rho} = 1 - \lim_{\ell^* \to 0}\left(\frac{\Delta p - \Delta p_0}{\ell^* \Delta p'}\right) + {\cal O}(\mu/\hbar\Omega_R)\,,
\label{eq:generalsignal}
\ee
where the corrections arise from the approximation
(\ref{eq:taylor}), which is accurate provided the atoms are in the
parabolic region of the lowest band.

In the limit of large $\Omega_R$,
the wavefunction of the lowest band is
$(\psi_{-1},\psi_0,\psi_1) = 1/2(1,-\sqrt{2},1)$
for all $\ell$.
Computing perturbative corrections to order $1/\Omega_R^2$ in $|\psi_m|^2$, we find
$\Delta p_0
\simeq
 (\delta/\Omega_R^2)\left[\sqrt{2}\Omega_R - \hbar(\Delta \ell)^2/(2MR^2) - 2\epsilon\right] +
 {\cal O}(1/\Omega_R^3)
$,
and
$\Delta p' = \sqrt{2}\hbar \Delta \ell/(MR^2\Omega_R) + {\cal O}(1/\Omega_R^2)$.
Combining this with (\ref{eq:lstar}) we obtain
\be
\frac{\rho_s}{\rho} = 1 - \lim_{\delta \to 0}\left(\frac{\Delta p - \Delta p_0}{2\hbar \delta (\Delta \ell)^2/(MR^2 \Omega_R^2)}\right)  + {\cal O}(1/\Omega_R)\,,
\label{eq:signal}
\ee
The limit $\omega \to 0$ in (\ref{eq:sfdefinition}), replaced here by $\omega_{\rm eff} \propto \delta \to 0$, is discussed further below.

Eqns.~(\ref{eq:generalsignal},~\ref{eq:signal}) show how a spectroscopic
measurement of the populations $N_m$ can lead to a direct measurement of the
superfluid fraction. That there is a connection between these 
quantities is a central result of this paper.

\begin{figure}[ttp]
\includegraphics[width=0.9\columnwidth]{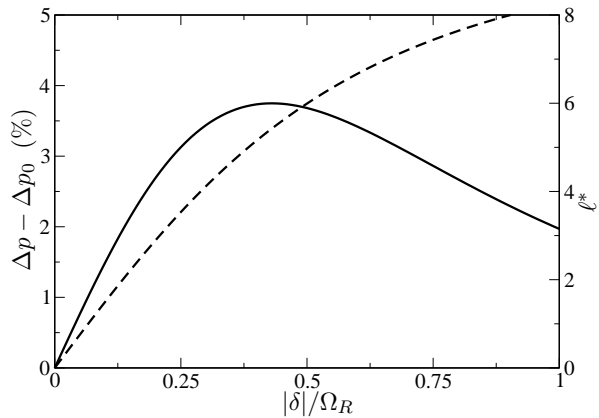}
\caption{ \label{fig:examples} Angular momentum
  $\ell^*$ at the bottom of the band (dashed line) and change in particle imbalance $\Delta p -\Delta p_0$ (solid line) as a function of
  $\delta/\Omega_R$ for a normal fluid (i.e. centred on $\ell^*$) for
  $\Omega_R = 1000 \hbar/M R^2$, $\Delta \ell = 10$, $\epsilon=0$.
This illustrates the precision required to distinguish a normal
fluid (here $\Delta p -\Delta p_0 \sim 3\%$) from a perfect superfluid
($\Delta p -\Delta p_0 = 0$).
}
\end{figure}

To distinguish a normal fluid from superfluid, the fractional population difference $\Delta p$ (\ref{eq:deltap}) must be measured with an absolute accuracy of order
\be
\frac{ 2 \hbar(\Delta \ell)^2\;\delta}{M R^2\;\Omega_R^2} \,.
\label{eq:sens}
\ee
This expression was
derived for $\delta/\Omega_R \ll 1$.  In
Fig.~\ref{fig:examples} we show the expected fractional change in
occupation for a {\it normal} fluid, with angular momentum centred on
$\ell^*$, computed for arbitrary $\delta/\Omega_R$.  This is shown for
parameters which for $^{23}$Na would correspond to $R=10 \, \mu \mbox{m}$,
$\Omega_R \simeq 2\pi\times 4.4 \, \mbox{kHz}$, and $\Delta \ell = 10$ (two
beams of orbital angular momentum 5).  In this case, $\Delta p$ must
be measured to an absolute accuracy of about $3\%$.
The required relative accuracy to distinguish a normal fluid ($\Delta p -\Delta p_0\neq 0$) from a
superfluid ($\Delta p = \Delta p_0$) is $(\Delta p - \Delta p_0)/\Delta p_0$,
and is about $10\%$  in the
linear regime $\delta/\Omega_R\lesssim 0.25$ in Fig.~\ref{fig:examples}.
This relatively small signal poses a moderate experimental challenge. It is
important to stress that it relies only on the measurement of {\it fractional}
occupations of different states. It is therefore insensitive to systematic
uncertainties in the absolute atom number determination,
and statistical errors can be reduced by
averaging over many shots. 

Above we used $\Delta \ell =10$ as a currently realistic experimental value, 
but the signal can be increased by increasing  $\Delta \ell$.
Since we require $\Omega_R \gtrsim \mu/\hbar$ 
for atoms to be in the parabolic region of the lowest band,
it is instructive to write (\ref{eq:sens}) in the form
$(\mu/\hbar\Omega_R)(4\delta/\Omega_R)(\Delta \ell \xi/R)^2$,
where $\xi$ is the healing
length. All three terms in this 
expression can in principle be close to unity.
One limitation to $\Delta \ell$ arises
from the fact that the expression (\ref{eq:signal}) applies in the
limit $\omega_{\rm eff} \propto \delta \to 0$, so that 
the imposed rotation is sufficiently small that the superfluid
does not change its angular momentum. This will fail if the resulting
velocity of the superfluid, $R \omega_{\rm
  eff} = \hbar \ell^*/(M^* R)$,
becomes larger than the superfluid critical velocity, which is
$\sim {\hbar}/({M\xi})$.  
If
this condition is violated, the superfluid will relax (vortices will
enter the system) and acquire
non-zero angular momentum.
The condition for
stability of the superfluid flow requires  $\Delta \ell$ to be less
than  $\sim (\Omega_R /\delta) R/\xi $ for $\delta \ll \Omega_R $
~(\ref{eq:lstar}),
and $\sim  R/\xi $ otherwise.
(Note that typically $\xi\leq 0.5 \, \mu$m.) 
As (\ref{eq:sens}) suggests, if
$\Delta\ell$ is limited by practical reasons, the signal will
generally be larger for lighter species, which for typical experimental parameters have lower density, and hence larger $\xi$.

Experimentally, a large spectroscopic signal which \emph{qualitatively}
distinguishes normal and superfluid components can be observed by a
``projective measurement": Suppose that the system is in equilibrium
in the parabolic band with $\Omega_R \gg \mu/\hbar \approx \delta$. If
we then reduce $\Omega_R$ to $\sim \delta$ on a time scale short
compared to the relaxation time but long compared to $1/\Omega_R$, the
superfluid and the normal fluids will remain centred at $\ell=0$ and
$\ell=\ell^*$, respectively, but the difference in their spin
composition will be greatly enhanced (see
Fig.~\ref{fig:curves}). However the \emph{quantitative} extraction of
the superfluid fraction would in this case require further analysis.

The ring-shaped trap discussed above is the case closest to the
theoretical discussions of the superfluid
fraction~\cite{leggettssleggettt0}, and the simplest to present.
However, our method also applies to a quasi-2D or 3D gas, provided the
optical fields are such that the atoms always remain in the lowest
energy band.
 As an illustration we consider a scenario in which two
hyperfine levels~\cite{spielmanpra}, labelled $\uparrow$ and
$\downarrow$, are coupled by the L-G beams propagating along $z$.  In
the $(\psi_\uparrow,\psi_\downarrow)$ basis, we parameterise the
lowest energy dressed eigenstate as $(e^{i\chi} \sin (\theta/2) ,
\cos(\theta/2))$, where $\theta(\bm{r})$ and $\chi(\bm{r})$ depend on
the local optical field.  For beams with angular momentum difference
$\Delta \ell$, we take $\chi({\bm r}) = \Delta \ell \phi$, where
$\phi$ is the azimuthal angle around the $z$ axis. Assuming optical
fields such that $\theta(\bm{r}) = \alpha r \ll 1$, where $r$ is the
radial distance in the $xy$ plane, the lowest energy state experiences
an effective vector potential that simulates uniform
rotation~\cite{blochdz}. The total number of flux quanta inside $r$ is
$\Delta \ell (\alpha r/2)^2$~ and $\omega_{\rm eff} =
(\hbar/4M)\Delta \ell \alpha^2$\cite{footnote}.  Computing the leading perturbative
corrections to the state, as in (\ref{eq:taylor}), one finds
$|\psi_\uparrow|^2 -|\psi_\downarrow|^2 \simeq \Delta p_0 + \Delta p'
r \langle p_\phi({\bm r}) \rangle$, where $\langle
p_\phi(\bm{r})\rangle$ is the local azimuthal momentum density. Thus,
the correction to $|\psi_\uparrow|^2 -|\psi_\downarrow|^2$ depends
linearly on the {\it local angular momentum density}.  The
spectroscopic measurement of $(N_\uparrow - N_\downarrow)/N$,
integrated over the sample, therefore provides a measure of the
angular momentum per particle $\langle L\rangle/N$.  Comparison of
$\langle L\rangle$ with $I_{\rm cl}\omega_{\rm eff}$ allows the
determination of the normal and superfluid fractions.

In summary, we have proposed a method to measure the superfluid
fraction of an ultracold atomic gas.  It combines the use of optical
beams with non-zero angular momentum to simulate rotation, with a
spectroscopic readout of angular momentum.  Our observation that
light-induced vector potentials create a direct connection between the
formal definition of superfluid density and the spin composition of a
gas is very general, and we expect it to be applicable to other
experimental scenarios.

\vskip0.1cm

\acknowledgments{We thank Jean Dalibard and  Mike Gunn for helpful comments. This work was supported by EPSRC Grant
  Nos. EP/F032773/1 (NRC) and EP/G026823/1 (ZH).}



\begin{thebibliography}{10}
\bibitem{footnote}
For stability of the superfluid flow, we require $\omega_{\rm eff} R  \lesssim \hbar/M\xi$ at the largest radius $R$.
\end{thebibliography}

\end{document}